\documentclass[12pt]{article}
\usepackage{amsfonts,amssymb,amsmath,mathrsfs,pict2e,hhline}
\usepackage{hyperref}
\usepackage[format=hang,labelsep=period,labelfont=bf,font=small]{caption}
\newcommand{\beq}{\begin{equation}}
\newcommand{\eeq}{\end{equation}}
\newcommand{\vp}{\vphantom}

\newcommand{\bs}{\boldsymbol}
\newcommand{\al}{\alpha}
\newcommand{\bt}{\beta}
\newcommand{\g}{\gamma}

\newcommand{\D}{\Delta}
\newcommand{\de}{\delta}
\newcommand{\varep}{\varepsilon}
\newcommand{\la}{\lambda}

\newcommand{\s}{\sigma}
\newcommand{\ta}{\tau}
\newcommand{\mc}{\mathcal}
\newcommand{\mr}{\mathrm}

\newcommand{\ads}{AdS_5\times S^5}
\newcommand{\coset}{PSU(2,2|4)/(SO(1,4)\times SO(5))}

\makeatletter
\DeclareRobustCommand{\loplus}{\mathbin{\mathpalette\dog@lsemi{+}}}
\DeclareRobustCommand{\lotimes}{\mathbin{\mathpalette\dog@lsemi{\times}}}
\DeclareRobustCommand{\roplus}{\mathbin{\mathpalette\dog@rsemi{+}}}
\DeclareRobustCommand{\rotimes}{\mathbin{\mathpalette\dog@rsemi{\times}}}

\newcommand{\dog@rsemi}[2]{\dog@semi{#1}{#2}{-90,90}}
\newcommand{\dog@lsemi}[2]{\dog@semi{#1}{#2}{270,90}}
\newcommand{\dog@semi}[3]{%
  \begingroup
  \sbox\z@{$\m@th#1#2$}%
  \setlength{\unitlength}{\dimexpr\ht\z@+\dp\z@\relax}%
  \makebox[\wd\z@]{\raisebox{-\dp\z@}{%
    \begin{picture}(1,1)
    \linethickness{\variable@rule{#1}}
    \roundcap
    \put(0.5,0.5){\makebox(0,0){\raisebox{\dp\z@}{$\m@th#1#2$}}}
    \put(0.5,0.5){\arc[#3]{0.5}}
    \end{picture}%
  }}%
  \endgroup
}
\newcommand{\variable@rule}[1]{%
  \fontdimen8
  \ifx#1\displaystyle\textfont3\else
    \ifx#1\textstyle\textfont3\else
      \ifx#1\scriptstyle\scriptfont3\else
        \scriptscriptfont3\relax
  \fi\fi\fi
}
\makeatother
\begin{document}
\begin{center}
\textbf{\Large Oscillator approach to quantization of \\[0.2cm] $AdS_5\times S^5$ superparticle in twistor formulation} \\[0.3cm]
{\large D.V.~Uvarov\footnote{E-mail: d\_uvarov@\,hotmail.com}}\\[0.2cm]
\textit{NSC Kharkov Institute of Physics and Technology,}\\ \textit{61108 Kharkov, Ukraine}\\[0.5cm]
\end{center}
\begin{abstract}
Using known relation between $SU(2,2|4)$ supertwistors and $SU(2)$ bosonic and fermionic oscillators we identify the physical states of quantized massless $\ads$ superparticle in supertwistor formulation and discuss how they fit into the spectrum of fluctuations of IIB supergravity on $\ads$ superbackground.
\end{abstract}
\section{Introduction}

Significant progress attained over the last decade and a half in perturbative $N=4$ super-Yang-Mills theory (see, e.g. \cite{Elvang} for review) was substantially triggered by the seminal work \cite{Witten'03} that, based on the previous findings \cite{PT'86}, \cite{Nair}, showed how string and twistor theories can be coupled in a synergetic way. One of the developments was the construction of (ambi)twistor action functionals for $N=4$ super-Yang-Mills theory in \cite{Mason'05}, \cite{MS}, \cite{Boels} that results in efficient rearrangement of the perturbative series expansion.

Since $N=4$ super-Yang-Mills theory has a dual description as Type IIB string theory on $\ads$ superbackground it is interesting to apply twistor methods also to the exploration of this duality \cite{Siegel}, \cite{Williams}. This necessitates working out supertwistor formulation on the $\ads$ superbackground for the superstring \cite{MT'98}, \cite{Rahmfeld} and Type IIB supergravity \cite{GM'85}, \cite{KRvN} that should start from identifying proper supertwistor variables. One of the feasible ways towards this goal is to consider supertwistor formulation on the $\ads$ superbackground of the massless superparticle model that arises in the infinite-tension limit of the superstring.

Such approach was pursued in \cite{Bars-ads5s5} (see also
\cite{Bars-twistor-string}) based on the generalization of
the extended superparticle model \cite{Bars-first-twistor} that has
extra dynamical variables to linearly realize
$SU(2,2|4)$ symmetry and extra gauge symmetries to get the same physical degrees of freedom as in the conventional superparticle models. In the model of Ref.~\cite{Bars-ads5s5}
appropriately fixing part of the gauge symmetries allows to
express the superparticle's Lagrangian in terms of the
supertwistors arranged into two $SU(2)$ doublets, one
of which is a straight-forward generalization of the
Ferber supertwistors \cite{Ferber}, while components of another
$SU(2)$ doublet of supertwistors have different Grassmann parity,
i.e. $SU(2,2)$ components
are Grassmann odd, whereas $SU(4)$
components are Grassmann even. Besides that supertwistors satisfy $SU(2,2|4)$-invariant quadratic constraints that are generators of the $su(2|2)$ gauge symmetry of the superparticle's action.

In \cite{U'18}, \cite{U'19} we have found that this supertwistor Lagrangian can be obtained also from the supertwistor formulation of the conventional massless superparticle model on $\ads$ superbackground \cite{MTT}, \cite{Bars'02}, \cite{Horigane}, \cite{Siegel-superparticle}, \cite{Heinze}. One of the benefits of this approach is that it yields the incidence relations connecting supertwistor components and $\ads$ superspace coordinates via the $\coset$ supercoset representative. Let us note that unlike the superspace formulation of the superparticle, in which the Lagrangian involves highly non-linear expressions for the $\ads$ supervielbein bosonic components, kinetic term and the constraints in the supertwistor formulation are quadratic facilitating Dirac quantization. In \cite{U'19} it was obtained the set of equations for the superparticle's wave function in the space of superambitwistors, however, open problem is to elaborate on the details of the Penrose transform in order to establish a mapping of the components of the wave function to the fields of Type IIB supergravity multiplet compactified on $\ads$ superbackground.

In this note we use close relationship \cite{Litov}, \cite{Ilieva}, \cite{Kallosh} of $SU(2,2|4)$ supertwistors and $SU(2)$ bosonic and fermionic oscillators that were used to construct positive energy unitary irreducible representations (uirs) of $PSU(2,2|4)$ \cite{BG'83}, \cite{GM'85}, to provide representation-theoretic characterization of the quantum states of $\ads$ superparticle. Similar analysis of the massive bosonic particle model on $AdS_5$ in the twistor formulation \cite{CRZ}\footnote{This twistor formulation of the bosonic particle on $AdS_5$ was recently revisited and generalized to dimensions 4 and 7 in \cite{ABGPKT}, for which one can benefit from using two-component spinors with complex, real and quaternionic entries respectively.} was performed in \cite{CKR}. After briefly recapitulating four-supertwistor formulation of the massless superparticle, we proceed to study oscillator realization of the $u(2,2|4)$ global symmetry and $su(2|2)$ gauge symmetry generators, analyze gauge-invariant subspace of the quantum states of the superparticle and show how they fit the fluctuation spectrum of IIB supergravity over the $\ads$ superbackground.

\section{Four-supertwistor formulation of massless superparticle on $\ads$ superbackground}

In \cite{U'18}, \cite{U'19} it was derived four-supertwistor representation of the Lagrangian of the massless $\ads$ superparticle starting from the first-order form of its superspace Lagrangian. Corresponding action functional reads
\beq\label{4-stwistor-lagr}
\begin{array}{rl}
S=&\int d\tau\mathscr L_{4-\mathrm{stwistor}} \\[0.2cm]
\mathscr L_{4-\mathrm{stwistor}}=&\frac{i}{2}\left(\bar{\mc Z}^i_{\mc A}\dot{\mc Z}^{\mc A}_i-\dot{\bar{\mc Z}}\vp{\bar{\mc Z}}^i_{\mc A}\mc Z^{\mc A}_i\right)+\frac{i}{2}(\bar\Psi^{q}_{\mc A}\dot\Psi^{\mc A}_{q}-\dot{\bar\Psi}^{q}_{\mc A}\Psi^{\mc A}_{q}) \\[0.2cm]
+&\Lambda^j{}_i(\bar{\mc Z}^i_{\mc A}\mc Z^{\mc A}_j-\frac12\de^i_j\bar{\mc Z}^k_{\mc A}\mc Z^{\mc A}_k)+\Lambda^p{}_q(\bar\Psi^{q}_{\mc A}\Psi^{\mc A}_{p}-\frac12\de^{q}_{p}\bar\Psi^{r}_{\mc A}\Psi^{\mc A}_{r}) \\[0.2cm]
+&\Lambda(\bar{\mc Z}_{\mc A}^{i}\mc Z^{\mc A}_{i}+\bar\Psi_{\mc A}^{q}\Psi^{\mc A}_{q})+i\Lambda^i_{q}\bar\Psi^{q}_{\mc A}\mc Z^{\mc A}_i+i\bar\Lambda^{q}_i\bar{\mc Z}^i_{\mc A}\Psi^{\mc A}_{q},
\end{array}
\eeq
where components of the two doublets of $SU(2,2|4)$ supertwistors have different Grassmann parity, that is Ferber or $c$-type supertwistors
\beq\label{c-type-def}
\mc Z^{\mc A}_{i}=\left(
\begin{array}{c}
Z^{\bs\al}_i \\ \eta^A_i
\end{array}
\right),\quad \bar{\mc Z}^i_{\mc A}=(\mc Z^{\mc B}_{i})^\dagger\mc H^{\mc B}{}_{\mc A}=(\bar Z^i_{\bs\al}\;\bar\eta^i_A),\quad
\mc H^{\mc B}{}_{\mc A}=\left(
\begin{array}{cc}
H^{\bs\bt}{}_{\bs\al} & 0 \\[0.2cm]
0 & \de^{B}_A
\end{array}
\right)
\eeq
have Grassmann-even $SU(2,2)$ components $Z^{\bs\al}_i$, $\bar Z^i_{\bs\al}$ and Grassmann-odd $SU(4)$ components $\eta^A_i$, $\bar\eta^i_A$, whereas $a$-type supertwistors
\beq\label{a-type-def}
\Psi^{\mc A}_{q}=\left(
\begin{array}{c}
\xi^{\bs\al}_q \\ L^A_q
\end{array}
\right),\quad \bar \Psi^q_{\mc A}=(\Psi^{\mc B}_{q})^\dagger\mc
H^{\mc B}{}_{\mc A}=(\bar\xi^q_{\bs\al}\;\bar L^q_A)
\eeq
have
Grassmann-odd $SU(2,2)$ components $\xi^{\bs\al}_q $,
$\bar\xi^q_{\bs\al}$ and Grassmann-even $SU(4)$ components
$L^A_q$, $\bar L^q_A$. The action (\ref{4-stwistor-lagr})
coincides with that obtained in Ref.~\cite{Bars-ads5s5} by partial
gauge fixing $2T$ superparticle model in $2+10$ dimensions.
The advantage of the approach pursued in \cite{U'18}, \cite{U'19} is not only in establishing
the direct connection with the superparticle's formulation in $\ads$
superspace but also in obtaining the incidence relations for the
components of supertwistors and superspace coordinates
parametrizing $\coset$ supercoset representative $\mc G$
\beq\label{c-supertwistor-incidence}
\mc Z^{\mc A}_i=\mc G^{\mc
A}{}_{\mc B}v^{\mc B}_i,\quad v^{\mc B}_i=\left(\begin{array}{c}
v^{\bs\bt}_i \\ 0 \end{array}\right),\quad\bar{\mc Z}^i_{\mc
A}=v^i_{\mc B}\mc G^{-1\,\mc B}{}_{\mc A}
\eeq
and
\beq\label{a-supertwistor-incidence} \Psi^{\mc A}_{q}=\mc G^{\mc
A}{}_{\mc B}\ell^{\mc B}_{q},\quad\ell_{q}^{\mc B}=\left(
\begin{array}{c}
0 \\[0.2cm] \ell_{q}^B
\end{array}\right),
\quad\bar\Psi^{q}_{\mc A}=\ell^{\, q}_{\mc B}\mc G^{-1\,\mc B}{}_{\mc A}.
\eeq
Unlike the eight-supertwistor formulation also considered in \cite{U'19}, in the four-supertwistor formulation there are only the first-class constraints introduced in (\ref{4-stwistor-lagr}) via the Lagrange multipliers -- seven bosonic
\begin{subequations}\label{bos-constr}
\begin{eqnarray}
& \bar{\mc Z}^i_{\mc A}\mc Z^{\mc A}_j-\frac12\de^i_j\bar{\mc Z}^k_{\mc A}\mc Z^{\mc A}_k\approx0, & \\[0.2cm]
& \bar\Psi^{q}_{\mc A}\Psi^{\mc A}_{p}-\frac12\de^{q}_{p}\bar\Psi^{r}_{\mc A}\Psi^{\mc A}_{r}\approx0, & \\[0.2cm]
& T=\bar{\mc Z}_{\mc A}^{i}\mc Z^{\mc A}_{i}+\bar\Psi_{\mc A}^{q}\Psi^{\mc A}_{q}\approx0 & \label{T-constr}
\end{eqnarray}
\end{subequations}
and eight fermionic
\beq\label{ferm-constr}
\bar\Psi^{q}_{\mc A}\mc Z^{\mc A}_i\approx0,\quad\bar{\mc Z}^i_{\mc A}\Psi^{\mc A}_{q}\approx0
\eeq
that are the generators of the classical $su(2|2)$ gauge symmetry of the superparticle's action. It was shown in \cite{U'18}, \cite{U'19} that, when this gauge symmetry is taken into account, the number of gauge-invariant physical degrees of freedom in the supertwistor formulation is the same as in the superspace formulation. The presence of only the first-class constraints greatly facilitates Dirac quantization of the model but requires insights into the Penrose transform of the functions of superambitwistors of $c$- and $a$-type as was discussed in \cite{U'19}, where the set of the first-order differential equations for the superparticle's wave function was derived.

\section{From $SU(2,2|4)$ supertwistors to $SU(2)$ oscillators}

Let us start this section with a few remarks regarding oscillator construction of the representations of $D=4$ conformal algebra and its $N=4$ supersymmetric extension. Bosonic oscillators have been utilized since 1960's in the field-theoretical models with underlying conformal symmetry to construct the generators of $so(2,4)\cong su(2,2)$ algebra and describe their quantum states, as well as to realize certain $su(2,2)$ representations (see, e.g. \cite{Nambu}, \cite{Barut} and notably \cite{MackTodorov}, where quantized twistors and oscillators were treated in the uniform way and used to describe ladder representations). In the mid 1980's the oscillator approach was advanced further in \cite{GSa}, \cite{BG'83} and applied in \cite{GM'85} to construct positive energy (lowest weight) unitary irreducible representations (uirs) of $su(2,2)$ (of which ladder representations constitute a particular case) that arise upon the compactification of Type IIB supergravity on $\ads$ superbackground. Subsequently the oscillator approach was also used to study the $AdS_5/CFT_4$ correspondence \cite{GMZ1}, \cite{GMZ2}.

Infinite-dimensional positive energy uirs, that can be constructed with the aid of quantized bosonic oscillators, have (half-)integer labels and correspond to finite-component fields in $AdS_5$ or on its conformal boundary given by the 4-dimensional Minkowski space-time. Their classification was finalized in \cite{Mack'77} based on the method of induced representations from the maximal parabolic subalgebra $(D\oplus sl(2,\mathbb C))\loplus K\subset su(2,2)$, where $D$ and $K$ are generators of the conformal dilatations and boosts.\footnote{Equivalence of these approaches was discussed in \cite{GMZ2} including a one-to-one mapping of the representation labels.}
As was shown in Ref.~\cite{Dobrev'07} positive energy uirs with (half-)integer labels  correspond to the holomorphic discrete series uirs of $su(2,2)$ or its limits. Recently there was proposed a generalization of the oscillator approach to describe uirs of $su(p,q)$ and its superextensions with non-(half-)integer labels \cite{GV}.

The construction of Ref.~\cite{GSa} makes use of the bosonic oscillators carrying fundamental representation(s) of the semisimple part of the maximal compact subalgebra of the non-compact Lie algebra in question. Upon generalization \cite{BG'83} to non-compact simple superalgebras they acquire superpartners -- fermionic oscillators transforming in the fundamental representation of the subalgebra of $R$-symmetry and generating its finite-dimensional uirs. For $psu(2,2|4)$ superalgebra (recall that $su(2,2|4)$ is not simple) this construction implies introduction of the two species of bosonic oscillators associated with two $su(2)$ subalgebras of the maximal compact subalgebra of $su(2,2)$ and two species of fermionic oscillators for two $su(2)$ subalgebras of $su(4)$. Positive energy uirs of $psu(2,2|4)$ are grouped into supermultiplets -- massive (long), massless (short) and doubleton supermultiplets -- generated from respective lowest-weight vectors (lwv) that carry certain finite-dimensional uirs of the maximal compact subalgebra of $su(2,2)$ and $su(2)\oplus su(2)\subset su(4)$. Each member of the supermultiplet corresponds to a finite-component field in $AdS_5$ or on its conformal boundary transforming according to some finite-dimensional representation of the $su(4)$ $R$-symmetry algebra. Classification of $psu(2,2|4)$ supermultiplets was done in \cite{DPetkova} without employing oscillators (see also \cite{GV} for consideration relying on oscillators).

Let us now present our definition \cite{U'16} of the bosonic $SU(2)$ oscillators. Introduce $SL(2,\mathbb C)$ constituents of the $SU(2,2)$ components of $c$-type supertwistors
\beq\label{bos-tw-comp}
Z^{\bs\al}_i=\left(
\begin{array}{c}
\mu^{\al}_i \\[0.2cm]
\bar\Lambda_{\dot\al i}
\end{array}
\right),\quad \bar Z^i_{\bs\al}=(\Lambda^i_\al\;\bar\mu^{\dot\al i}).
\eeq
Then two species of the $SU(2)$ bosonic oscillators are defined by the linear combinations of the twistor components (\ref{bos-tw-comp})
\beq\label{def-su22-bos-osc1}
\left(\begin{array}{c}
a^{\al}_i \\
b^\al_i
\end{array}\right)=
\frac{1}{\sqrt{2}}\left(
\begin{array}{cc}
-I & I \\
I & I
\end{array}\right)\,
\left(\begin{array}{c}
\mu^{\al}_i \\[0.2cm]
\bar\Lambda_{\dot\al i}
\end{array}
\right)
=\frac{1}{\sqrt{2}}\left(
\begin{array}{c}
-\mu^\al_i+\bar\Lambda_{\dot\al i} \\
\mu^\al_i+\bar\Lambda_{\dot\al i}
\end{array}\right)
\eeq
and
\beq\label{def-su22-bos-osc2}
(a_\al^i\hspace{0.2cm}b^i_{\al})=
\frac{1}{\sqrt{2}}
(\bar\mu^{\dot\al i}\; \Lambda^i_\al)\,
\left(\begin{array}{cc}
-I & I \\
I & I
\end{array}\right)=\frac{1}{\sqrt{2}}(-\bar\mu^{\dot\al i}+\Lambda^i_\al\hspace{0.4cm}\bar\mu^{\dot\al i}+\Lambda^i_\al),
\eeq
where $I$ is the unit $2\times2$ matrix.
Unitary transformation (\ref{def-su22-bos-osc1})-(\ref{def-su22-bos-osc2}) describes transition from the basis of $D=2+4$ $\gamma$-matrices, in which the generator of conformal dilatations $D$ and $
so(1,3)$ generators are block-diagonal in the spinor representation, to the basis, in which generators of the maximal compact subalgebra $\mc K=su(2)\oplus su(2)\oplus u(1)\subset so(2,4)$ take block-diagonal form. The generators of the $su(2)$ algerbas in $\mc K$ are given by the linear combinations of the $su(2)\subset so(1,3)$ generators and  $so(2,4)-so(1,3)$ generators that have no time indices (see \cite{GMZ2} and Appendix B in \cite{GV} for details). The former basis corresponds to the realization of the $su(2,2)$ algebra as $D=4$ conformal algebra that is at the heart of the twistor theory, while the latter underlies the oscillator approach. Transformation (\ref{def-su22-bos-osc1})-(\ref{def-su22-bos-osc2})  amounts to the diagonalization of the matrix $H^{\bs\al}{}_{\bs\bt}$ that enters the definition of dual supertwistors in (\ref{c-type-def}) and (\ref{a-type-def})
\beq
H^{\bs\al}{}_{\bs\bt}=\left(\begin{array}{cc}
0 & I \\
I & 0
\end{array}
\right)\quad\rightarrow\quad
\left(\begin{array}{cc}
-I & 0 \\
0 & I
\end{array}
\right).
\eeq
Note some abuse of the notation in (\ref{def-su22-bos-osc1})-(\ref{def-su22-bos-osc2}) and below since, as in \cite{GV}, small lightface Greek letters stand for the indices of both $sl(2,\mathbb C)$ two-component spinors and two-component spinors of the two $su(2)$ algerbas in $\mc K$. Moreover, following \cite{GSa} we use the same letters to label indices of $a$- and $b$-oscillators although they carry spinor representations of different $su(2)$ algerbas and are not contracted. It suffices to take into account that $sl(2,\mathbb C)$ spinor indices are carried by the $SU(2,2)$ twistor components and space-time fields (discussed at the end of Section 5), whereas oscillators and $psu(2,2|4)$ supersymmetry generators constructed out of them carry $su(2)$ spinor indices.

Dirac bracket (D.B.) relations for the bosonic components of $c$-type supertwistors that follow from the kinetic term of the Lagrangian (\ref{4-stwistor-lagr})
\beq
\{Z^{\bs\al}_i,\bar Z^j_{\bs\bt}\}_{D.B.}=i\de^j_i\de^{\bs\al}_{\bs\bt}
\eeq
in the quantum theory become commutators\footnote{We do not place hats over the quantum operators not to overburden the notation. Hopefully this will not cause a confusion.}
\beq
[Z^{\bs\al}_i,\bar Z^j_{\bs\bt}]=\de^j_i\de^{\bs\al}_{\bs\bt}
\eeq
that upon above transformation translate into commutators of $a$- and $b$-oscillators
\beq
[a_\al^i,a^\bt_j]=\de^i_j\de_\al^\bt,\quad
[b^\al_i,b_\bt^j]=\de^j_i\de^\al_\bt.
\eeq
Thus oscillators $a^{\al}_i$ and $b^i_\al$ are treated as raising operators, whereas $a^i_\al$ and $b^\al_i$ are lowering operators annihilating oscillator vacuum $|0\rangle$.

Similarly D.B. relations of the odd components of $c$-type supertwistors (\ref{c-type-def}) read
\beq
\{\eta^A_i,\bar\eta^j_B\}_{D.B.}=i\de^j_i\de^A_B.
\eeq
On the quantum level they become anticommutators
\beq\label{su4-fermi-osc-anticomm}
\{\eta^A_i,\bar\eta^j_B\}=\de^j_i\de^A_B
\eeq
that can be viewed as defining relations of the two sets of $SU(4)$ fermionic oscillators. Decomposing $SU(4)$ (anti)fundamental representation on the fundamental representations of its two $SU(2)$ subgroups introduces two kinds of $SU(2)$ fermionic oscillators \cite{GM'85}, \cite{GMZ1}\footnote{Another option -- to use $SU(4)$ oscillators as they stand to construct $psu(2,2|4)$ uirs was considered in \cite{Bars0205}, \cite{Marboe}, \cite{GV}.}
\beq\label{def-su4-fermi-osc}
\eta^A_i=\left(
\begin{array}{c}
\al^a_i \\[0.2cm] \bt^{\dot a}_i
\end{array}
\right),\qquad
\bar\eta^i_A=\left(
\al^i_a \;
\bt^i_{\dot a}
\right)
\eeq
that satisfy
\beq
\{\al_i^a,\al^j_b\}=\de^j_i\de^a_b,\quad\{\bt^{\dot a}_i,\bt^j_{\dot b}\}=\de_i^j\de^{\dot a}_{\dot b}.
\eeq
Admissible choice is to treat $\al^a_i$ and $\bt^i_{\dot a}$ as raising oscillators, then $\al^i_a$ and $\bt^{\dot a}_i$ are lowering ones annihilating oscillator vacuum $|0\rangle$.

Note that in such a way we get two copies of bosonic and fermionic oscillators of each kind transforming as the doublet of one of the $su(2)$ subalgebras of $su(2|2)$ gauge superalgebra that is an instance of the color algebras in the oscillator approach. Note also that to realize $su(2,2)$ and $psu(2,2|4)$ generators it suffices to take just one copy of the oscillators of each kind. Using these minimal realization of $su(2,2)$ it is possible to construct only ladder uirs (named doubletons in the oscillator approach) and their supermultiplets, among which the $N=4$ super-Yang-Mills multiplet is the only self-conjugate one \cite{GM'85}, \cite{GMZ1}. Other positive energy $su(2,2)$ (and $psu(2,2|4)$) uirs can be obtained by decomposing tensor products of finitely many (super)doubletons. 

One can analogously introduce oscillators associated with the $a$-type supertwistors (\ref{a-type-def}). Namely, taking linear combinations of the $SL(2,\mathbb C)$ constituents of the odd components of $a$-type supertwistors
\beq
\xi^{\bs\al}_q=\left(
\begin{array}{c}
m^\al_q \\ \bar n_{\dot\al q}
\end{array}
\right),\quad\bar\xi^q_{\bs\al}=(n^q_\al\;\bar m^{\dot\al q})
\eeq
defines another two species of $SU(2)$ fermionic oscillators
\beq\label{def-su22-fermi-osc}
\begin{array}{c}
\omega^\al_q=\frac{1}{\sqrt{2}}(-m^\al_q+\bar n_{\dot\al q}),\quad \omega_\al^q=\frac{1}{\sqrt{2}}(-\bar m^{\dot\al q}+n^q_\al),\\[0.2cm]
\varphi^\al_q=\frac{1}{\sqrt{2}}(m^\al_q+\bar n_{\dot\al q}),\quad \varphi^q_{\al}=\frac{1}{\sqrt{2}}(\bar m^{\dot\al q}+n^q_\al)
\end{array}
\eeq
in addition to (\ref{def-su4-fermi-osc}). The fact that Grassmann parity of the $SU(2,2)$ components of $a$-type supertwistors differs from that of the $c$-type supertwistors has far-reaching consequences for the future analysis. In particular, from the anticommutators of $\xi^{\bs\al}_q$ and $\bar\xi^p_{\bs\bt}$
\beq
\{\xi^{\bs\al}_q,\bar\xi^p_{\bs\bt}\}=\de^p_q\de^{\bs\al}_{\bs\bt}
\eeq
it follows that
\beq\label{anticom-omega-phi}
\{\omega^\al_q,\omega_\bt^p\}=-\de^p_q\de^\al_\bt,\quad\{\varphi^\al_q,\varphi_\bt^p\}=\de^p_q\de^\al_\bt,
\eeq
i.e. $\omega$-oscillators satisfy wrong-sign anticommutation relations.

Grassmann-even components of $a$-type supertwistors that in the quantum theory satisfy commutation relations
\beq\label{su4-even-comm-rel}
[L^A_q,\bar L^p_B]=\de^p_q\de^A_B
\eeq
admit decomposition on the $SU(2)$ constituents in analogy with (\ref{def-su4-fermi-osc})
\beq\label{def-su4-bos-osc}
L^A_q=\left(
\begin{array}{c}
c^a_q \\[0.2cm] d^{\dot a}_q
\end{array}
\right),\quad\bar L^q_A=(c^q_a\; d^q_{\dot a}).
\eeq
This yields two extra species of $SU(2)$ bosonic oscillators. Commutation relations for them can be read off from (\ref{su4-even-comm-rel})
\beq\label{a-type-bos-osc-comm}
[c^a_q,c^p_b]=\de^p_q\de^a_b,\quad [d^{\dot a}_q,d^p_{\dot b}]=\de^p_q\de^{\dot a}_{\dot b}.
\eeq
As will be shown in the next section for one type of the introduced of $SU(2)$ bosonic oscillators the commutation relations are necessarily the wrong-sign ones.

\section{Oscillator realization of global $u(2,2|4)$ symmetry}

As was mentioned above superparticle's action (\ref{4-stwistor-lagr}) is invariant under the $su(2|2)$ gauge symmetry. The constraints (\ref{bos-constr}) and (\ref{ferm-constr}) span the $su(2|2)$ superalgebra named color superalgebra in \cite{Bars-twistor-string}, \cite{Bars-ads5s5} since it generalizes bosonic color symmetries that accompany  oscillator realizations of $su(2,2)$ and $psu(2,2|4)$ generators \cite{GM'85}, \cite{GMZ1}, \cite{GMZ2}, \cite{Bars0205}. Superparticle's action (\ref{4-stwistor-lagr}) is also invariant under $U(2,2|4)$ global symmetry generated by
\beq\label{u224-generators}
\mc Z^{\mc A}_i\bar{\mc Z}^i_{\mc B}-\Psi^{\mc A}_q\bar\Psi^q_{\mc B}=u(2,2|4).
\eeq
Important property of these $u(2,2|4)$ generators is that they (anti)commute with the generators of the $su(2|2)$ gauge symmetry due to the color symmetry invariance.\footnote{This is a manifestation of the Howe duality that has been recently discussed in \cite{Karapet} with the emphasis on the oscillator approach.} When expressed in terms of oscillators introduced in (\ref{def-su22-bos-osc1}), (\ref{def-su22-bos-osc2}), (\ref{def-su4-fermi-osc}), (\ref{def-su22-fermi-osc}) and (\ref{def-su4-bos-osc}), these global symmetry generators are used to build the space of states of quantized superparticle model as the certain representation space of $u(2,2|4)$. These generators are of three types: quadratic in creation oscillators, quadratic in annihilation oscillators and composed of the products of creation and annihilation oscillators in accordance with the three-grading structure of the $(ps)u(2,2|4)$ superalgebra \cite{GMZ1}. This three-grading structure allows to choose which of the oscillators associated with the $a$-type supertwistors represent creation and annihilation operators once such choice is made for the oscillators related to the $c$-type supertwistors.

Consider first the $su(2,2)\subset u(2,2|4)$ algebra. Three-grading decomposition of its generators is as follows
\beq\label{3-grading-su22}
g_{(0)}=\{E,L^\al{}_\bt,R^\al{}_\bt\},\quad g_{(-1)}=\{T^{(-)\,\al}{}_{\bt}\},\quad g_{(+1)}=\{T^{(+)\,\al}{}_{\bt}\}:
\eeq
\vspace*{-1cm}
\begin{subequations}
\begin{eqnarray}
& 2E=a^\al_i a^i_\al+b_\al^i b^\al_i-\omega^\al_q\omega_\al^q+\varphi_\al^q\varphi^\al_q, & \label{ads-energy} \\
& L^\al{}_\bt=a^\al_ia_\bt^i-\omega_q^\al\omega^q_\bt-\frac12\de^\al_\bt(a_i^\g a^i_\g-\omega_q^\g\omega^q_\g), & \label{su2-l} \\
& R^{\,\al}{}_\bt=b^\al_ib_\bt^i-\varphi^\al_q\varphi_\bt^q-\frac12\de^\al_\bt(b_i^\g b^i_\g-\varphi_q^\g\varphi^q_\g), & \label{su2-r} \\
& T^{(-)\,\al}{}_{\bt}=b^\al_ia_\bt^i-\varphi_q^\al\omega^q_\bt, & \label{su22-grade-1} \\
& T^{(+)\,\al}{}_{\bt}=a_i^\al b_\bt^i-\omega^\al_q\varphi^q_\bt. & \label{su22-grade+1}
\end{eqnarray}
\end{subequations}
Above expressions deserve a number of comments. From (\ref{su22-grade-1}) and (\ref{su22-grade+1}) it follows that $\omega^q_\bt$ and $\varphi_q^\al$ are annihilation operators, whereas $\omega^\al_q$ and $\varphi^q_\bt$ are creation operators. In (\ref{3-grading-su22}) $g_{(0)}$ is spanned by the generators of the maximal compact subalgebra $u(1)_E\oplus su(2)\oplus su(2)$ of $su(2,2)$, where $E$ is the $AdS_5$ energy (conformal dimension) operator that defines grading of $su(2,2)$ generators:
\beq
[E,g_{(\s)}]=\s g_{(\s)}\quad\s=0,\pm1.
\eeq
As a result $su(2,2)$ commutation relations can be written in concise schematic form
\beq
[g_{(\s)},g_{(\s')}]=g_{(\s+\s')}.
\eeq
This three grading decomposition can be viewed as a modification of the Cartan decomposition.
Expression (\ref{ads-energy}) corresponds to creation-annihilation ordering of oscillators and was obtained from the manifestly Hermitian one by moving raising oscillators to the left. Introducing oscillator number operators $N_{(a)}=a^\al_i a^i_\al$, $N_{(b)}=b_\al^i b^\al_i$, $N_{(\omega)}=-\omega^\al_q\omega_\al^q$ and $N_{(\varphi)}=\varphi_\al^q\varphi^\al_q$ it can be brought to the form
\beq\label{E-expressed-in-osci-numbers}
E=\frac12(N_{(a)}+N_{(b)}+N_{(\omega)}+N_{(\varphi)}).
\eeq
Note that $N_{(\omega)}$ includes extra minus sign that correlates with the wrong sign in the anticommutator of the $\omega$-oscillators in (\ref{anticom-omega-phi}). Though $E$ is defined to be non-negative, in (\ref{E-expressed-in-osci-numbers}) there is no positive $c$-number term in distinction to the case when $su(2,2)$ generators are realized solely in terms of the bosonic oscillators \cite{GM'85}, \cite{GMZ1}. Such positive $c$-number contribution ensures that obtained irreducible representations (irreps) lie above or saturate respective unitarity bounds. There is no ordering ambiguity in the expressions (\ref{su2-l}) and (\ref{su2-r}) for $L^\al{}_\bt$ and $R^{\,\al}{}_\bt$ that generate two $su(2)$ subalgebras of $su(2,2)$. lwv's carry spin $s_1$ and $s_2$ irreps of these $su(2)$ subalgebras and are eigenvectors of $E$ so that two spins $s_1$, $s_2$ and $AdS_5$ energy are, as is common, used to label the $su(2,2)$ irreps. Whatever oscillator realization of the $su(2,2)$ generators is, positive energy representations are constructed by repeated action of $T^{(+)\,\al}{}_{\bt}$ generators on the lwv's annihilated by $T^{(-)\,\al}{}_{\bt}$. In the model under consideration lwv corresponding to the superparticle's ground state is also annihilated by the $su(2|2)$ gauge symmetry generators.

Generators of the $su(4)\subset u(2,2|4)$ algebra also admit three-grading decomposition
\beq\label{3-grading-su4}
h_{(0)}=\{C,A^a{}_b,B^{\dot a}{}_{\dot b}\},\quad h_{(-1)}=\{V^{(-)\,\dot a}{}_{b}\},\quad h_{(+1)}=\{V^{(+)\, a}{}_{\dot b}\}:
\eeq
\vspace*{-1cm}
\begin{subequations}
\begin{eqnarray}
& 2C=-\al_i^a\al^i_a-\bt^i_{\dot a}\bt^{\dot a}_i+c_q^ac^q_a-d^q_{\dot a}d^{\dot a}_q, & \\
& A^a{}_b=\al_i^a\al^i_b-c_q^ac^q_b-\frac12\de^a_b(\al_i^d\al^i_d-c_q^dc^q_d), & \\
& B^{\,\dot a}{}_{\dot b}=\bt^{\dot a}_i\bt^i_{\dot b}-d^{\dot a}_qd^q_{\dot b}-\frac12\de^{\dot a}_{\dot b}(\bt^{\dot c}_i\bt^i_{\dot c}-d^{\dot c}_qd^q_{\dot c}), & \\
& V^{(-)\,\dot a}{}_{b}=\bt^{\dot a}_i\al^i_b-d_q^{\dot a}c^q_b, & \label{su4-grade-1} \\
& V^{(+)\, a}{}_{\dot b}=\al^a_i\bt^i_{\dot b}-c_q^ad^q_{\dot b} &  \label{su4-grade+1}
\end{eqnarray}
\end{subequations}
so that commutation relations of $su(4)$ algebra can be
brought to the form respecting this three-grading structure
\beq
[h_{(\ta)},h_{(\ta')}]=h_{(\ta+\ta')},
\eeq
where grading of a
generator is read off from its commutator with $C$
\beq
[C,h_{(\ta)}]=-\ta h_{(\ta)}\quad\ta=0,\pm1.
\eeq
The form of the
diannihilation and dicreation operators (\ref{su4-grade-1}) and
(\ref{su4-grade+1}) suggests that bosonic oscillators $c^q_b$ and
$d_q^{\dot a}$ are lowering ones, while $c_q^a$ and $d^q_{\dot b}$
are raising. In view of (\ref{a-type-bos-osc-comm}) this choice
implies that $c$-oscillators satisfy wrong-sign commutation
relations. Then the operator $C$ can be written in the form
similar to that of the $AdS_5$ energy operator
(\ref{E-expressed-in-osci-numbers})
\beq\label{C-expressed-in-osci-numbers}
C=-\frac12(N_{(\al)}+N_{(\bt)}+N_{(c)}+N_{(d)}),
\eeq
where
$N_{(\al)}=\al_i^a\al^i_a$, $N_{(\bt)}=\bt^i_{\dot a}\bt^{\dot
a}_i$, $N_{(c)}=-c_q^ac^q_a$ and $N_{(d)}=d^q_{\dot a}d^{\dot
a}_q$ are oscillator number operators and the overall minus sign has
been introduced to conform with the definition of $C$ in
\cite{GMZ1}. It is readily seen to be non-positive definite,
whereas for the $su(4)$ uirs constructed in \cite{GM'85} its eigenvalues are
non-negative due to the positive $c$-number contribution equal the
number of copies of the oscillators. $su(4)$ irreps are labelled by two spins $j_{1}$ and $j_{2}$ of the irreps of $su(2)$ algebras spanned by $A^a{}_b$ and $B^{\,\dot a}{}_{\dot b}$ that are carried by lwv's as well as eigenvalues of $C$. Like $su(2,2)$ lowest-weight representations, $su(4)$ lowest-weight representations are constructed via repeated action of $V^{(+)\, a}{}_{\dot b}$ on respective lwv's
annihilated by $V^{(-)\,\dot a}{}_{b}$. If $su(4)$ generators are constructed using finite number of fermionic oscillators as in \cite{GM'85}, \cite{GMZ1} resulting lowest-weight representations are finite-dimensional.

Two additional bosonic generators of $u(2,2|4)$ are the constraint (\ref{T-constr}) that in oscillator form equals
\beq
T=-N_{(a)}+N_{(b)}-N_{(\al)}+N_{(\bt)}-N_{(\omega)}+N_{(\varphi)}-N_{(c)}+N_{(d)}\approx0
\eeq
and
\beq
U=-N_{(a)}+N_{(b)}+N_{(\al)}-N_{(\bt)}-N_{(\omega)}+N_{(\varphi)}+N_{(c)}-N_{(d)}.
\eeq
Since $T$ is the constraint, for all physical states of the quantized superparticle its eigenvalue is zero but eigenvalues of $Y=-N_{(a)}+N_{(b)}-N_{(\omega)}+N_{(\varphi)}$ can be used as an additional label together with the $su(2,2)$ and $su(4)$ labels.

Among 32 odd generators of $u(2,2|4)$ eight are constructed out of the raising oscillators
\beq\label{susy-raising}
Q^{(+)\,\al}{}_{\dot a}=a^\al_i\bt^i_{\dot a}-\omega^\al_q d^q_{\dot a},\quad Q^{(+)\, a}{}_\al=\al^a_i b^i_\al-c_q^a\varphi^q_\al
\eeq
and belong to the $+1$ eigenstate w.r.t. the grading defined by $E-C$. Their repeated action on the $su(2,2)\oplus su(4)$ lwv of the ground state of the superparticle, that will be identified in the next Section, yields all the lwv's of the supermultiplet of the physical states.

\section{Physical states of quantized superparticle}

As discussed above $u(2,2|4)$ generators (\ref{u224-generators}) (anti)commute with $su(2|2)$ gauge symmetry generators (\ref{bos-constr}) and (\ref{ferm-constr}), so the latter act directly on the lwv's and one can examine the constraints they impose on the ground state of the quantized superparticle.  Bosonic constraints (\ref{bos-constr}) imply that sought for lwv is a singlet of the two $su(2)$ color subalgebras of $su(2|2)$ and is annihilated by $T$. So the ground state can be constructed by acting on the vacuum $|0\rangle$ with the products of the color-singlet combinations of raising oscillators, namely, the determinants of bosonic oscillators $\D_{(a)}=\frac12\varep^{ij}\varep_{\al\bt}a^\al_i a^\bt_j$, $\D_{(b)}=\frac12\varep_{ij}\varep^{\al\bt}b_\al^i b_\bt^j$, $\D_{(c)}=\frac12\varep^{pq}\varep_{ab}c^a_p c^b_q$ and $\D_{(d)}=\frac12\varep_{pq}\varep^{\dot a\dot b}d^p_{\dot a} d^q_{\dot b}$, and antisymmetrized in color indices products of the fermionic oscillators $\al^{ab}=\al^a_i\varep^{ij}\al^b_j$, $\bt_{\dot a\dot b}=\bt^i_{\dot a}\varep_{ij}\bt^j_{\dot b}$, $\omega^{\al\bt}=\omega^\al_p\varep^{pq}\omega^\bt_q$ and $\varphi_{\al\bt}=\varphi^p_\al\varep_{pq}\varphi^q_\bt$ raised to the powers balanced in such a way as to be annihilated by $T^{(-)\,\al}{}_{\bt}$, $V^{(-)\,\dot a}{}_{b}$ and $T$. Among these lwv's only vacuum itself is annihilated by all the fermionic constraints (\ref{ferm-constr}) and hence is the ground state.

Other lwv's of the supermultiplet of physical states are obtained by repeated action of the supersymmetry generators (\ref{susy-raising}) on the ground state. Consider first lwv's that can be obtained by acting solely with $Q^{(+)\,\al}{}_{\dot a}$ generators. For example, the product of two such supersymmetry generators can be decomposed on three types of contributions depending on the oscillators involved
\beq
Q^{(+)\,\al_1}{}_{\dot a_1}Q^{(+)\,\al_2}{}_{\dot a_2}=a^{\al_1}_ia^{\al_2}_j\bt^i_{\dot a_1}\bt^j_{\dot a_2}+\omega^{\al_1}_q\omega^{\al_2}_p d^q_{\dot a_1}d^p_{\dot a_2}+a^{\al_1}_i\omega^{\al_2}_q\bt^i_{\dot a_1}d^q_{\dot a_2}-a^{\al_2}_i\omega^{\al_1}_q\bt^i_{\dot a_2}d^q_{\dot a_1}.
\eeq
Further decomposing the products of oscillators on irreducible tensors gives two lwv's:
\beq
-\frac14\varep_{\dot a_1\dot a_2}\left(a^{(\al_1}_ia^{\al_2)}_j\bt^{ij}-2\omega^{\al_1\al_2}\D_{(d)}+2a^{(\al_1}_i\omega^{\al_2)}_q\bt^i_{\dot b}\varep^{\dot b\dot c}d^q_{\dot c}\right)|0\rangle,\quad\bt^{ij}=\bt^i_{\dot b}\varep^{\dot b\dot c}\bt^j_{\dot c}
\eeq
with spins $s_1=1$, $s_2=0$ with respect to $su(2)$ subalgebras of $su(2,2)$ and $j_1=j_2=0$ with respect to $su(2)$ subalgebras of $su(4)$, and
\beq
\frac14\varep^{\al_1\al_2}\left(2\D_{(a)}\bt_{\dot a_1\dot a_2}-\omega_{pq}d^p_{(\dot a_1}d^q_{\dot a_2)}-2a^\bt_i\varep_{\bt\g}\omega^\g_p\bt^i_{(\dot a_1}d^p_{\dot a_2)}\right)|0\rangle,\quad\omega_{pq}=\omega^{\bt}_p\varep_{\bt\g}\omega^\g_q
\eeq
with spins $s_1=s_2=0$ and $j_1=0$, $j_2=1$ presented in the third and fourth rows of the Table \ref{abeta-lwv}. Since each of the other products of $Q^{(+)\,\al}{}_{\dot a}$'s gives rise to single lwv and the products of more than four supercharges vanish, one finds six independent lwv's presented in the Table \ref{abeta-lwv}.
\begin{table}
\renewcommand{\arraystretch}{1.4}
\begin{center}
{\footnotesize
\begin{tabular}{|c|c|c|c|}
\hline
$su(2,2)$ labels $(E,s_1,s_2)$ & $su(4)$ labels $(C,j_1,j_2)$ & $Y$ & fields on $\ads$ \\
\hline \hline
$(0,0,0)$ & $(0,0,0)$ & 0 & $\varphi^{(1)}(y,z)$ \\
\hline
$(1/2,1/2,0)$ & $(-1/2,0,1/2)$ & -1 & $\la^{(1)}_{\al\dot a}(y,z)$ \\
\hline
$(1,1,0)$ & $(-1,0,0)$ & -2 & $A^{(1)}_{\al(2)}(y,z)$ \\
\hline
$(1,0,0)$ & $(-1,0,1)$ & -2 & $\varphi^{(2)}_{\dot a(2)}(y,z)$ \\
\hline
$(3/2,1/2,0)$ & $(-3/2,0,1/2)$ & -3 & $\la^{(2)}_{\al\dot a}(y,z)$ \\
\hline
$(2,0,0)$ & $(-2,0,0)$ & -4 & $\varphi^{(3)}(y,z)$ \\
\hline
\end{tabular}
\caption{$su(2,2)$ and $su(4)$ labels of lwv's obtained by acting with the powers of $Q^{(+)\,\al}{}_{\dot a}$ on $|0\rangle$ and respective fields on $\ads$}\label{abeta-lwv}
}
\end{center}
\end{table}
Spin content of these lwv's coincides with the spin content of the lwv's in the first column of the Table 1 in \cite{GM'85}\footnote{We prefer to label lwv's by four spins of four $su(2)$ subalgebras of $psu(2,2|4)$ rather than by Young tableau as in \cite{GM'85}.} that have units in the second and third digits. We normalize $AdS_5$ energy $E$ in such a way that its eigenvalues are (half-)integer for lwv's with (half-)integer $s_1+s_2$ and eigenvalues of $Y$ take integer values.

Similarly repeated application of $Q^{(+)\, a}{}_\al$ produces five other lwv's with interchanged spins $s_1\leftrightarrow s_2$ and $j_1\leftrightarrow j_2$ presented in the Table \ref{balpha-lwv}.
\begin{table}
\renewcommand{\arraystretch}{1.4}
\begin{center}
{\footnotesize
\begin{tabular}{|c|c|c|c|}
\hline
$su(2,2)$ labels $(E,s_1,s_2)$ & $su(4)$ labels $(C,j_1,j_2)$ & $Y$ & fields on $\ads$ \\
\hline \hline
$(1/2,0,1/2)$ & $(-1/2,1/2,0)$ & 1 & $\la^{(1)}_{\dot\al a}(y,z)$ \\
\hline
$(1,0,1)$ & $(-1,0,0)$ & 2 & $ A^{(1)}_{\dot\al(2)}(y,z)$ \\
\hline
$(1,0,0)$ & $(-1,1,0)$ & 2 & $\varphi^{(2)}_{a(2)}(y,z)$ \\
\hline
$(3/2,0,1/2)$ & $(-3/2,1/2,0)$ & 3 & $\la^{(2)}_{\dot\al a}(y,z)$ \\
\hline
$(2,0,0)$ & $(-2,0,0)$ & 4 & $\bar\varphi^{(3)}(y,z)$ \\
\hline
\end{tabular}
\caption{$su(2,2)$ and $su(4)$ labels of lwv's obtained by acting with the powers of $Q^{(+)\, a}{}_\al$ on $|0\rangle$ and respective fields on $\ads$}\label{balpha-lwv}
}
\end{center}
\end{table}
Their spin content is the same as that of the lwv's given in the first column of the Table 1 in
\cite{GM'85} with units in the
first and fourth digits. Multiple application of the bosonic generators (\ref{su22-grade+1}) and (\ref{su4-grade+1}) to obtained lwv's produces basis vectors of the infinite-dimensional $su(2,2)\oplus su(4)$ irreps. Respective fields on $\ads$, coordinatized by $y$ and $z$, are listed in the utmost right column of the Tables \ref{abeta-lwv} and \ref{balpha-lwv}. To make contact with the consideration of Ref.~\cite{GM'85} we use similar notation for these fields. In particular, fields with the same spins of $su(2)\oplus su(2)\subset su(2,2)$ are denoted by the same letters and superscripts with numbers in round brackets are used to label them in order of increasing $AdS_5$ energy. In the subscripts numbers in round brackets following $sl(2,\mathbb C)$ or $su(2)$ indices denote a group of symmetrized indices equal to that number.

There remains to consider simultaneous application
of $Q^{(+)\,\al}{}_{\dot a}$ and $Q^{(+)\, a}{}_\al$ to the ground state. To
find respective lwv's one has to decompose the products of oscillators
into irreps of four $su(2)$ subalgebras of $psu(2,2|4)$ and leave,
according to the criterion formulated in \cite{Bars0205}, only
traceless tensors of the two $su(2)$ color subalgebras. The role of the latter requirement can be explained on the simple example of the product $Q^{(+)\,\al}{}_{\dot a}Q^{(+)\, b}{}_\bt$. There arise four contributions
\beq
Q^{(+)\,\al}{}_{\dot a}Q^{(+)\, b}{}_\bt=a^\al_i b^j_\bt\bt^i_{\dot a}\al^b_j+\omega^\al_p\varphi^q_\bt d^p_{\dot a}c^b_q+a^\al_i\varphi^q_\bt\bt^i_{\dot a}c^b_q-\omega^\al_p b^j_\bt d^p_{\dot a}\al^b_j.
\eeq
It is easy to see that neither $a^\al_i b^j_\bt|0\rangle$ and $\omega^\al_p\varphi^q_\bt|0\rangle$ nor their traces in the color indices form $su(2,2)$ lwv's since the r.h.s. of the commutators of $a^\al_i b^j_\bt$ and $\omega^\al_p\varphi^q_\bt$ with the lowering generators (\ref{su22-grade-1}) do not annihilate the oscillator vacuum. However, color-traceless combinations $(a^\al_i b^j_\bt-1/2\de^j_ia^\al_k b^k_\bt)|0\rangle$ and $(\omega^\al_p\varphi^q_\bt-1/2\de^q_p\omega^\al_r\varphi^r_\bt)|0\rangle$ indeed are the lwv's, as well as  $a^\al_i\varphi^q_\bt|0\rangle$ and $\omega^\al_p b^j_\bt|0\rangle$ that ab initio transform irreducibly under both color subalgebras. Note that due to the color symmetry invariance of the $su(2,2)\oplus su(4)$ lwv's, taking color-traceless combinations of the products of oscillators associated with $su(2,2)$ selects color-traceless combinations of the products of oscillators associated with $su(4)$. In fact there is a relationship between the $su(2)\oplus su(2)\subset su(2,2)$ representations, $su(2)\oplus su(2)\subset su(4)$ representations and those of the color symmetry carried by a lwv.\footnote{In the non-supersymmetric case color symmetry is not essential in constructing $su(2,2)$ lwv's (see, e.g. sections 5-6 in Ref.~\cite{GMZ1}).} So $(\bt^i_{\dot a}\al^b_j-1/2\de^i_j\bt^k_{\dot a}\al^b_k)|0\rangle$ and $(d^p_{\dot a}c^b_q-1/2\de^p_qd^r_{\dot a}c^b_r)|0\rangle$ form $su(4)$ lwv's together with $\bt^i_{\dot a}c^b_q|0\rangle$ and $d^p_{\dot a}\al^b_j|0\rangle$. Respective spin values $s_1=s_2=1/2$ and $j_1=j_2=1/2$ are given in the first row of the Table \ref{mixed}.
Examining other products of the supercharges yields extra 24 lwv's, whose labels can formally be obtained by summing those of all pairs of the lwv's from Tables \ref{abeta-lwv} and
\ref{balpha-lwv}, except for the vacuum. 
\begin{table}
\renewcommand{\arraystretch}{1.4}
\begin{center}
{\footnotesize
\begin{tabular}{|c|c|c|c|}
\hline
$su(2,2)$ labels $(E,s_1,s_2)$ & $su(4)$ labels $(C,j_1,j_2)$ & $Y$ & fields on $\ads$ \\
\hline \hline
$(1,1/2,1/2)$ & $(-1,1/2,1/2)$ & 0 & $A^{(1)}_{\al\dot\al a\dot a}(y,z)$ \\
\hline
$(3/2,1,1/2)$ & $(-3/2,1/2,0)$ & -1 & $\Psi^{(1)}_{\al(2)\dot\al a}(y,z)$ \\
\hline
$(3/2,0,1/2)$ & $(-3/2,1/2,1)$ & -1 & $\la^{(3)}_{\dot\al a\dot a(2)}(y,z)$ \\
\hline
$(3/2,1/2,1)$ & $(-3/2,0,1/2)$ & 1 & $\Psi^{(1)}_{\al\dot\al(2)\dot a}(y,z)$ \\
\hline
$(3/2,1/2,0)$ & $(-3/2,1,1/2)$ & 1 & $\la^{(3)}_{\al a(2)\dot a}(y,z)$ \\
\hline
$(2,1,0)$ & $(-2,1,0)$ & 0 & $A^{(2)}_{\al(2)a(2)}(y,z)$ \\
\hline
$(2,0,1)$ & $(-2,0,1)$ & 0 & $A^{(2)}_{\dot\al(2)\dot a(2)}(y,z)$ \\
\hline
$(2,1,1)$ & $(-2,0,0)$ & 0 & $h_{\al(2)\dot\al(2)}(y,z)$ \\
\hline
$(2,0,0)$ & $(-2,1,1)$ & 0 & $\varphi^{(4)}_{a(2)\dot a(2)}(y,z)$ \\
\hline
$(2,1/2,1/2)$ & $(-2,1/2,1/2)$ & -2 & $A^{(2)}_{\al\dot\al a\dot a}(y,z)$ \\
\hline
$(2,1/2,1/2)$ & $(-2,1/2,1/2)$ & 2 & $\bar A^{(2)}_{\al\dot\al a\dot a}(y,z)$ \\
\hline
$(5/2,1/2,1)$ & $(-5/2,0,1/2)$ & -1 & $\Psi^{(2)}_{\al\dot\al(2)\dot a}(y,z)$ \\
\hline
$(5/2,1/2,0)$ & $(-5/2,1,1/2)$ & -1 & $\la^{(5)}_{\al a(2)\dot a}(y,z)$ \\
\hline
$(5/2,1,1/2)$ & $(-5/2,1/2,0)$ & 1 & $\Psi^{(2)}_{\al(2)\dot\al a}(y,z)$ \\
\hline
$(5/2,0,1/2)$ & $(-5/2,1/2,1)$ & 1 & $\la^{(5)}_{\dot\al a\dot a(2)}(y,z)$ \\
\hline
$(5/2,0,1/2)$ & $(-5/2,1/2,0)$ & -3 & $\la^{(4)}_{\dot\al a}(y,z)$ \\
\hline
$(5/2,1/2,0)$ & $(-5/2,0,1/2)$ & 3 & $\la^{(4)}_{\al\dot a}(y,z)$ \\
\hline
$(3,1/2,1/2)$ & $(-3,1/2,1/2)$ & 0 & $A^{(3)}_{\al\dot\al a\dot a}(y,z)$ \\
\hline
$(3,0,1)$ & $(-3,0,0)$ & -2 & $ A^{(3)}_{\dot\al(2)}(y,z)$ \\
\hline
$(3,0,0)$ & $(-3,1,0)$ & -2 & $\varphi^{(5)}_{a(2)}(y,z)$ \\
\hline
$(3,1,0)$ & $(-3,0,0)$ & 2 & $A^{(3)}_{\al(2)}(y,z)$ \\
\hline
$(3,0,0)$ & $(-3,0,1)$ & 2 & $\varphi^{(5)}_{\dot a(2)}(y,z)$ \\
\hline
$(7/2,0,1/2)$ & $(-7/2,1/2,0)$ & -1 & $\la^{(6)}_{\dot\al a}(y,z)$ \\
\hline
$(7/2,1/2,0)$ & $(-7/2,0,1/2)$ & 1 & $\la^{(6)}_{\al\dot a}(y,z)$ \\
\hline
$(4,0,0)$ & $(-4,0,0)$ & 0 & $\varphi^{(6)}(y,z)$ \\
\hline
\end{tabular}
\caption{$su(2,2)$ and $su(4)$ labels of lwv's obtained by acting with the powers of $Q^{(+)\,\al}{}_{\dot a}$ and $Q^{(+)\, a}{}_\al$ on $|0\rangle$ and respective fields on $\ads$}\label{mixed}
}
\end{center}
\end{table}

We conclude this section with a few comments regarding comparison of the results obtained in this note with known descriptions of the compactification spectrum of IIB supergravity on $\ads$ superbackground. In distinction to the approach of Ref.~\cite{GM'85} superparticle model in the supertwistor formulation has finitely many physical states. This can be explained by the fact that in \cite{GM'85} the spectrum of IIB supergravity on $\ads$ is described in terms of the $S^5$ Fourier modes of the on-shell $10d$ fields, that is the fields on $AdS_5$ transforming according to finite-dimensional $su(4)$ uirs. Fourier modes presented in each line of the Table 1 in \cite{GM'85} have the same $su(2)\oplus su(2)\subset su(2,2)$ spins and are parametrized by integer $p$ with $AdS_5$ energies and $su(4)$ labels proportional to $p$.
The states of the superparticle model collected in Tables 1-3 (apart from the leading mode of $\varphi^{(1)}$ that corresponds to the scalar field carrying the $su(2,2)$ representation) are naturally identified with the $\ads$ fields obtained by summing Fourier modes in each line of the Table 1 in \cite{GM'85}. On the other hand, Fourier modes with the same $p$ in different lines make up supermultiplets of the $psu(2,2|4)$ supersymmetry.  In particular, leading modes in the $S^5$ harmonic expansion of $\la^{(1)}_{\al\dot a}$, $\la^{(1)}_{\dot\al a}$, $A^{(1)}_{\al(2)}$, $A^{(1)}_{\dot\al(2)}$ and subleading mode of $\varphi^{(1)}$ make up $N=4$ super-Yang-Mills multiplet that corresponds to modes with $p=1$ in Table 1 of \cite{GM'85}. This multiplet is known to decouple from other modes given by the $D=5$ $N=8$ gauged supergravity multiplet ($p=2$) and massive $psu(2,2|4)$ supermultiplets with spins ranging from 0 to 2 ($p>2$).

$AdS_5$ energies of some of the superparticle's states in Tables 1-3 with spins $s_1$ and $s_2$ lie below the unitarity bound for massless fields of $AdS_5$ $E\geq s_1+s_2+2$ ($s_1s_2\not=0$) and the bound for massless fields on the Minkowski boundary of $AdS_5$ $E\geq s_1+s_2+1$ ($s_1s_2=0$)\footnote{Unitarity conditions found by G.~Mack \cite{Mack'77} for the $SU(2,2)$ positive energy irreps  are the special cases of the unitarity bounds for the positive energy irreps of $so(2,d)$ \cite{Metsaev} and for the positive energy irreps of $su(p,q)$ \cite{GV}.} and hence correspond to non-unitary representations with non-negative energy. To explain the possible origin of such irreps take, for instance, fields $\varphi^{(1)}$ in the first row of the Table 1 in \cite{GM'85}.
Respective $su(2,2)$ uirs have ehergies $E^{\,\varphi^{(1)}}_p=p$ (in our normalization) and their sum $\sum\limits^{\infty}_{p=1}E^{\,\varphi^{(1)}}_p$  diverges but can be $\zeta-$regularized to take finite value of $-1/12$. Also the description of the $su(4)$ irreps by means of ever increasing number of fermionic oscillators in the limit results in the infinite-dimensional non-unitary irreps that can be realized in terms of finitely many bosonic oscillators. Making this arguments more rigorous requires further study.

Finally let us confront the results of the present work with those of Ref.~\cite{KRvN}, where the spectrum of IIB supergravity on $\ads$ superbackground had been obtained starting from the field equations in $D=1+9$ dimensions expanded over the background values of the metric and Ramond-Ramond (RR) self-dual five-form and linearized in small excitations. This allowed not only to find the set of fields on $AdS_5$ but also the equations they satisfy and to trace their $10d$ origin. So Table \ref{origin} below serves twofold purpose -- to match the fields that arise in the spectrum of physical states of the considered superparticle model with those found in \cite{KRvN} and to establish their relation to the fields from the $D=1+9$ supermultiplet of IIB supergravity. 
\begin{table}
\renewcommand{\arraystretch}{1.4}
\begin{center}
{\footnotesize
\begin{tabular}{|c|c|c|@{}p{0.1cm}@{}|c|c|c|}
\hline
\multicolumn{3}{|c|}{Bosonic fields} & &\multicolumn{3}{|c|}{Fermionic fields} \\
\cline{1-3} \cline{5-7}
 & \multicolumn{2}{c|}{$\ads$ fields} & & & \multicolumn{2}{c|}{$\ads$ fields}  \\
\cline{2-3} \cline{6-7}
\raisebox{2.2ex}[0cm][0cm]{$10d$ fields} & Ref.~\cite{KRvN} & ours & & \raisebox{2.2ex}[0cm][0cm]{$10d$ fields} & Ref.~\cite{KRvN} & ours \\
\cline{1-3} \cline{5-7}
& $h'_{\mu\nu}$ & $h_{\al(2)\dot\al(2)}$ & & & & \\
\cline{2-3}
&  & $A^{(1)}_{\al\dot\al a\dot a}$ & & & & \raisebox{2.5ex}[0cm][0cm]{$\Psi^{(1)}_{\al(2)\dot\al a}$} \\
& \raisebox{1.7ex}[0cm][0cm]{$h_{\mu\al}$} & $A^{(3)}_{\al\dot\al a\dot a}$ & & & \raisebox{4.5ex}[0cm][0cm]{$\varphi^{\mr L+}_{(\mu)}$} & \raisebox{1.5ex}[0cm][0cm]{$\Psi^{(1)}_{\al\dot\al(2)\dot a}$} \\
\cline{2-3} \cline{6-7}
\raisebox{2ex}[0cm][0cm]{graviton} & $h_{(\al\bt)}$ & $\varphi^{(4)}_{a(2)\dot a(2)}$ & & & & $\Psi^{(2)}_{\al(2)\dot\al a}$ \\
\cline{2-3}
& & $\varphi^{(1)}$ & & & \raisebox{1.9ex}[0cm][0cm]{$\varphi^{\mr L-}_{(\mu)}$} & $\Psi^{(2)}_{\al\dot\al(2)\dot a}$ \\ \cline{6-7}
& \raisebox{1.5ex}[0cm][0cm]{$h^\al_\al$} & $\varphi^{(6)}$ & & complex & & $\la^{(1)}_{\al\dot a}$ \\
\cline{1-3}
complex scalar & $B$ & $\varphi^{(3)},\,\bar\varphi^{(3)}$ & & gravitino & \raisebox{1.7ex}[0cm][0cm]{$\Psi^{\mr L-}$} & $\la^{(1)}_{\dot\al a}$  \\
\cline{1-3} \cline{6-7}
 & & $A^{(1)}_{\al(2)}$, $A^{(1)}_{\dot\al(2)}$ & & & & $\la^{(6)}_{\al\dot a}$ \\
& \raisebox{1.7ex}[0cm][0cm]{$A_{\mu\nu}$} & $A^{(3)}_{\al(2)}$, $A^{(3)}_{\dot\al(2)}$ & & & \raisebox{1.7ex}[0cm][0cm]{$\Psi^{\mr L+}$} & $\la^{(6)}_{\dot\al a}$ \\
\cline{2-3} \cline{6-7}
complex 2-form & $A_{\mu\al}$ & $A^{(2)}_{\al\dot\al a\dot a}$, $\bar A^{(2)}_{\al\dot\al a\dot a}$ & & & & $\la^{(3)}_{\al a(2)\dot a}$ \\
\cline{2-3}
& & $\varphi^{(2)}_{a(2)}$, $\varphi^{(2)}_{\dot a(2)}$ & & & \raisebox{1.7ex}[0cm][0cm]{$\Psi^{\mr T-}$} & $\la^{(3)}_{\dot\al a\dot a(2)}$ \\
\cline{6-7}
& \raisebox{1.7ex}[0cm][0cm]{$A_{\al\bt}$} & $\varphi^{(5)}_{a(2)}$, $\varphi^{(5)}_{\dot a(2)}$ & & & & $\la^{(5)}_{\al a(2)\dot a}$ \\
\cline{1-3}
 & $a_{\mu\nu\al\bt}$ & $A^{(2)}_{\al(2)a(2)}$, $A^{(2)}_{\dot\al(2)\dot a(2)}$ & & & \raisebox{1.7ex}[0cm][0cm]{$\Psi^{\mr T+}$} & $\la^{(5)}_{\dot\al a\dot a(2)}$ \\
\cline{2-3} \cline{5-7}
& & $A^{(1)}_{\al\dot\al a\dot a}$ & & & & $\la^{(2)}_{\al\dot a}$  \\
RR 4-form & \raisebox{1.7ex}[0cm][0cm]{$a_{\mu\al\bt\g}$} & $A^{(3)}_{\al\dot\al a\dot a}$ & & complex & \raisebox{1.7ex}[0cm][0cm]{$\la^{\mr L-}$} & $\la^{(2)}_{\dot\al a}$ \\
\cline{2-3} \cline{6-7}
& & $\varphi^{(1)}$ & & spinor & & $\la^{(4)}_{\al\dot a}$ \\
& \raisebox{1.7ex}[0cm][0cm]{$a_{\al\bt\g\de}$} & $\varphi^{(6)}$ & & & \raisebox{1.7ex}[0cm][0cm]{$\la^{\mr L+}$} & $\la^{(4)}_{\al\dot a}$ \\
\hline
\end{tabular}
\caption{Correspondence between the fields of IIB supegravity and the spectrum of its compactification on $\ads$ superbackground}\label{origin}
}
\end{center}
\end{table}
The data presented require some explanatory comments. In distinction to our notation, in \cite{KRvN} small Greek letter are used for curved-space indices of $AdS_5$ ($\mu,\nu=0,\ldots,4$) and $S^5$ ($\al,\bt=1,\ldots,5$), whereas indices of four-components spinors of $Spin(1,4)$ and $Spin(5)$ are not shown explicitly. Bosonic fields that appear twice in the third column, such as $\varphi^{(1)}$ and $\varphi^{(6)}$, stand for eigenvectors of the $2\times2$ matrices that enter the systems of coupled equations for the fields in the second column, like $S^5$ Fourier modes of $h^\al_\al$ and $a_{\al\bt\g\de}$. Homogeneous equations for $A_{\mu\nu}$ factorize so that there are two solutions and $S^5$ expansion of $A_{\al\bt}$ includes harmonics of two types that are eigenfunctions of the 'square root' of the $S^5$ Laplacian with different eigenvalues also resulting in two independent solutions. Expansions of the fermionic fields include spinor (labelled by $\mr L$) and vector-spinor (labelled by $\mr T$) harmonics that are eigenspinors of the Dirac operator on $S^5$ with opposite eigenvalues so expansion modes are endowed $+$ and $-$ superscripts. The presence of spinor and vector-spinor harmonics with different eigenvalues results in two independent solutions with different 'masses' ($AdS_5$ energies) of the equations for fermionic fields of each type.

\section{Discussion}

In this note we have shown how physical states of the quantized
$\ads$ massless superparticle in the twistor formulation can be
described in terms of the $su(2)$ bosonic and fermionic oscillators
some of which were used previously to construct unitary supermultiplets of
$psu(2,2|4)$ \cite{GM'85}, \cite{GMZ1}, \cite{GMZ2}. These physical
states of the superparticle have been mapped to those in the spectrum of fluctuations of IIB supergravity on $\ads$ superbackground
\cite{GM'85}, \cite{KRvN} and $N=4$ super-Yang-Mills multiplet on $D=4$ Minkowski boundary of $AdS_5$.  We observed that the description of
IIB supergravity spectrum on $\ads$ superbackground in the context of $D=10$ superparticle model differs
from the original one in Ref.~\cite{GM'85}, where it was
presented in the form of an infinite sum of unitary $psu(2,2|4)$ supermultiplets,
lwv's of which were obtained by acting on the oscillator vacuum with
raising supersymmetry generators of the $psu(2,2|4)$ superalgebra constructed out of $P\geq1$ copies of the four sets of $su(2)$ oscillators
coming from the linear combinations of the components of $SU(2,2|4)$ Ferber supertwistors. In
these supermultiplets fields corresponding to the irreps that have the same $su(2)\oplus su(2)\subset su(2,2)$
spins and $Y$ eigenvalues but transform according to different
representations of $su(4)$ represent
expansion coefficients of the $10d$ fields over $S^5$ harmonics \cite{KRvN}. On the contrary, dynamical variables of the $\ads$
superparticle model in the four-supertwistor formulation are two
Ferber or $c$-type supertwistors and two $a$-type supertwistors,
whose components have another Grassmann parity. The latter give
rise to $su(2)$ oscillators some of which satisfy wrong-sign
(anti)commutation relations. As a result for some states their $su(2,2)$ lwv's have $AdS_5$ energies,
though non-negative but lying below the unitarity bounds for respective values of
spins and hence corresponding to non-unitary irreps. We conjectured that such irreps correspond to infinite sums of the unitary ones.  There
also arise infinite-dimensional non-unitary $su(4)$
representations that can be interpreted as resulting from summing up the
$S^5$ Fourier modes.

It would be interesting to study quantization of the superparticle model directly in terms of the supertwistors. In \cite{U'19}
there were derived the first-order differential equations for
the superparticle's wave function, whose arguments are $c$- and
$a-$type superambitwistors. To provide the supertwistor description of the IIB supergravity fluctuations over $\ads$ superbackground as solution of these
equations it is necessary to develop ambitwistor transform of respective fields on $AdS_5$. This would be an important step towards the twistor formulation of the $AdS_5/CFT_4$ duality.

The approach reported here can also be extended to describe fluctuation spectra over other highly supersymmetric backgrounds of known supergravity theories in the framework of appropriate supertwistor formulations of the models of superparticles and extended objects such as superstrings and branes.

The author is grateful to A.A.~Zheltukhin for interesting discussions. This work was partially supported by the Ministry of Education and Science of Ukraine (research grant No.0120U102252).

\end{document}